\documentstyle[prb,aps,multicol,epsf]{revtex}
\epsfclipon
\begin {document}
\title{Self-generated magnetic flux in YBa$_2$Cu$_3$O$_{7-x}$ grain
boundaries}
\author{R.~G.~Mints and Ilya~Papiashvili}
\address{School of Physics and Astronomy,\\
Raymond and Beverly Sackler Faculty of Exact Sciences, Tel Aviv
University, Tel Aviv 69978, Israel}
\date{\today}
\maketitle
\begin{abstract}
Grain boundaries in YBa$_2$Cu$_3$O$_{7-x}$ superconducting films are
considered as Josephson junctions with a critical current density
$j_c(x)$ alternating along the junction. A self-generated magnetic
flux is treated both analytically and numerically for an almost
periodic distribution of $j_c(x)$. We obtained a magnetic flux-pattern
similar to the one which was recently observed experimentally.
\end{abstract}
\begin{multicols}{2}
\narrowtext
%
%%%%%%%%%% INTRODUCTION %%%%%%%%%%%
%
\section{Introduction}
Grain boundaries in high-$T_c$ cuprates are interesting and important
for both fundamental physics and applications of high-temperature
superconductivity.\cite{Lin,Tsu,Mil} Conventional models of strongly
coupled Josephson junctions are applicable to describe electromagnetic
properties of grain boundaries in thin films of high-$T_c$
superconductors.\cite{Cha,Gro} A remarkable exception of this rule is
the [001]-tilt boundary in YBa$_2$Cu$_3$O$_{7-x}$ films with a
misorientation angle close to 45$^{\circ}$.\cite{Iva,Che,Hum,Cop,Man}
Indeed, these grain boundaries have an anomalous dependence of the
critical current $I_c$ on an applied magnetic field $H_a$. Contrary to
a usual Fraunhofer-type dependence $I_c(H_a)$ with a major central peak
at $H_a=0$ and minor symmetric side-peaks the asymmetric 45$^{\circ}$
[001]-tilt grain boundaries demonstrate a pattern without a central
dominant peak. Instead two symmetric major side-peaks appear at certain
applied magnetic fields $H_a=\pm\, H_{\rm sp}\ne
0$.\cite{Iva,Che,Hum,Cop,Man,Min}
\par
Several mechanisms have been suggested to explain this phenomena.
\cite{Hum,Cop,Min} The anomalous dependence $I_c(H_a)$ with symmetric
major side-peaks is obviously a result of a specific heterogeneity of
electrical properties of the asymmetric 45$^{\circ}$ [001]-tilt grain
boundaries. Two fundamental experimental observations in conjunction
explain this heterogeneity in a natural way.\cite{Cop,Man} {\it First},
a fine scale faceting of grain boundaries was discovered in experiments
using the transmission electron microscopy (TEM) technique.
\cite{Mil,Jia,Ros,Tra,Seo} These facets have a typical length-scale $l$
of the order of 10--100 nm and a wide variety of orientations relative
to the axis of symmetry of the superconductor. {\it Second}, quite a
few of recent experiments provide an evidence of a predominant
$d_{x^2-y^2}$ wave symmetry of the order parameter in many of the
high-$T_c$ cuprates. In some experimental studies the symmetry of the
order parameter is more complicated and is shown to be a certain
mixture of the $d_{x^2-y^2}$ and $s$ wave
components.\cite{Tsu,DJM,Wol,Bra,Igu,Van,Ish,Mul}
\par
These two fundamental experimental observations indicate the existence
of two contributions to the phase difference of the order parameter
across the grain boundary. Indeed, consider a meandering grain boundary
in a film of a superconductor with the $d_{x^2-y^2}$ wave symmetry of
the order parameter and assume that there is a certain magnetic flux
inside the grain boundary. In this case there is a phase difference
$\varphi$ caused by the magnetic flux and at the same time there is an
additional phase difference  $\alpha$ caused by the misalignment of the
anisotropic $d_{x^2-y^2}$ wave superconductors. Therefore, the
tunneling current density $j_c$ is defined by the total phase
difference $\Delta =\varphi -\alpha$. A model describing this Josephson
current density $j(x)$ results from an assumption that
$j(x)\propto\sin\,[\varphi (x)-\alpha (x)]$, where $x$ is along the
grain boundary line.\cite{Man} The local values of the phase difference
$\alpha (x)$ depend on the relative orientation of neighboring facets.
In the case of an asymmetric 45$^{\circ}$ grain boundary we have
$\alpha (x)=0$ or $\pi$, and therefore $j(x)\propto\sin\varphi
(x)\cos\alpha (x)$.\cite{Man} In other words in the framework of a
model relating $j(x)$ to the orientation of the facets we arrive to
$j(x)=j_c(x)\sin\varphi (x)$ with an alternating critical current
density $j_c(x)\propto\cos\alpha(x)$. The dependence $j_c(x)$ is
imposed by the sequence of facets along a grain boundary line. If this
sequence is periodic or almost periodic then the function $j_c(x)$ is a
periodic or almost periodic alternating function. The typical
length-scale for $j_c(x)$ is of the same order as the length of the
facets $l$, {\it i.e.}, this length-scale is of 10--100 nm.
\par
Variation of orientation of facets along a meandering grain boundary
leads to formation of local superconducting current loops even in the
absence of an applied magnetic field if the total phase difference
$\Delta\ne 0$. It was predicted, in particular, that these current
loops can generate a certain magnetic flux at a contact of two facets
with $\alpha =0$ and $\alpha =\pi$.\cite{Cop}
\par
Self-generated randomly distributed magnetic flux was discovered in
asymmetric 45$^{\circ}$ [001]-tilt grain boundaries in
YBa$_2$Cu$_3$O$_{7-x}$ superconducting films in the absence of an
applied external magnetic field.\cite{JMa} This flux $\phi_s(x)$
changes its sign randomly and has an amplitude of variations less than
the flux quantum $\phi_0$. The average value of $\phi_s(x)$ along the
grain boundary is nearly zero.  Noticeable, that this disordered
non-quantized magnetic flux was observed only for the samples
exhibiting the anomalous dependence of the critical current $I_c(H_a)$
on magnetic field with the two symmetric major side-peaks.
\par
It was shown analytically that under certain conditions a stationary
state with a self-generated flux exists for a Josephson junction with
a periodically alternating critical current density
$j_c(x)$.\cite{RGM} The same spatial distributions of $j_c(x)$ result
in an anomalous dependence of the critical current $I_c(H_a)$ on the
applied magnetic field.\cite{Min} Numerical calculations show that two
symmetric major side-peaks appear for a periodically alternating
$j_c(x)$. The randomness of the critical current density $j_c(x)$
smears these peaks but leaves their position in place at weak
randomness. We therefore conclude that the experimental observation of
the well pronounced major side-peaks on the curve $I_c(H_a)$
\cite{Iva,Che,Hum,Cop,Man} means that the alternating critical current
density is a periodic or almost periodic function of $x$. A noticeable
randomness of $j_c(x)$ would smear out the dominant
side-peaks.\cite{Min}
\par
In this paper we calculate both analytically and numerically the
self-generated flux $\phi_s$ in a Josephson junction with an almost
periodically alternating critical current density $j_c(x)$. The paper
is organized as follows. First, we review briefly the case when
$j_c(x)$ is a periodic alternating function.\cite{RGM} We derive the
main equations of the two-scale perturbation theory and apply these
equations to analyze the non-quantized self-generated flux. This
approach forms a basis for the following analytical calculations. Next,
we treat the self-generated flux for the case of an almost periodic
alternating critical density $j_c(x)$ and start with a qualitative
approach to the problem. We review then the results of our numerical
simulations which verify the qualitative consideration and exhibit a
magnetic flux-pattern which is similar to the one that was recently
observed experimentally.\cite{JMa}
\par
%
%%%%%%%%%% Main equations %%%%%%%%%%
%
\section{Main equations}
It is convenient for the following analyses to write the function
$j_c(x)$ in the form
% #1
\begin{equation}
j_c=\langle j_c\rangle\,[1+g(x)],
\label{eq1}
\end{equation}
where $\langle j_c\rangle$ is the average value of the critical
current density $j_c(x)$ over an interval with a length $L\gg l$
% #2
\begin{equation}
\langle j_c\rangle ={1\over L}\,\int_0^Lj_c(x)\,dx.
\label{eq2}
\end{equation}
\par
The function $g(x)$ introduced in Eq.~(\ref{eq1}) alternates on a
typical length scale of $l$. Note that by definition the average value
of $g(x)$ is zero, {\it i.e.}, $\langle g(x)\rangle =0$. The maximum
value of $|g(x)|$ varies from $|g(x)|_{\rm max}\gtrsim 1$ to
$|g(x)|_{\rm max}\gg 1$. We assume also that $\lambda\ll
l\ll\Lambda_J$, where $\lambda$ is the London penetration depth and
% #3
\begin{equation}
\Lambda_J^2={c\,\phi_o\over 16\pi^2\lambda\langle j_c\rangle}
\label{eq3}
\end{equation}
is an effective Josephson penetration depth. It is worth to mention
that in the case of an alternating current density this effective
penetration depth is not a local characteristics of a tunnel junction.
It is rather an effective quantity defined on the same typical length
scale as $\langle j_c\rangle$.
\par
The phase difference $\varphi(x)$ satisfies the equation
% #4
\begin{equation}
\Lambda_J^2\,\varphi''-[1+g(x)]\sin\varphi=0.
\label{eq4}
\end{equation}
In the limiting case $l\ll\Lambda_J$ it is convenient to write a
solution of this equation as a sum of a certain smooth function
$\psi(x)$ with a length scale of order $\Lambda_J$ and a rapidly
oscillating function $\xi(x)$ with a length scale of order
$\l$\cite{RGM}
% #5
\begin{equation}
\varphi(x)=\psi(x)+\xi(x).
\label{eq5}
\end{equation}
We assume also that $|\xi(x)|\ll|\psi(x)|$. After substituting
Eq.~(\ref{eq5}) into Eq.~(\ref{eq4}) and keeping the terms up to the
first order in $\xi (x)$ we obtain
% #6
\begin{equation}
\Lambda_J^2\psi ''+\Lambda_J^2\xi '' -
[1+g(x)][\sin\psi+\xi\cos\psi]=0.
\label{eq6}
\end{equation}
\par
Note, that experimentally\cite{JMa} the self-generated flux was
observed by a SQUID pickup loop with a size of several
$\Lambda_J\gg~l$. It means that this method is averaging out the fast
alternating flux defined by the phase $\xi (x)$ and measures the
spatially smooth flux defined by $\psi (x)$.
\par
Next we consider briefly the case of a {\it periodically} alternating
critical current density $j_c(x)$ which forms the basis of the
following analysis of a general case with $j_c(x)$ being a randomly
alternating function.
\par
%
%%%%%%%%%%Periodically alternating critical current density%%%%%%%%%%%
%
\section{Periodically alternating critical current density}
\subsection{Two-scale perturbation theory}
If the critical current density $j_c(x)$ is a periodic function, then
$g(x)$ also is a periodic function. In this case an approximate
solution of Eq.~(\ref{eq6}) can be obtained based on a two-scale
perturbation theory.\cite{Lan} As a first step in order to apply this
approach to Eq.~(\ref{eq6}) we separate the fast alternating terms
with a typical length scale $l$ and the smooth terms varying with a
typical length scale $\Lambda_J$
% #7
\begin{equation}
(\Lambda_J^2\psi''-\sin\psi -g\xi\cos\psi )+
(\Lambda_J^2\xi'' -g\sin\psi)=0.
\label{eq7}
\end{equation}
\par
In Eq.~(\ref{eq7}) we omitted two out of three fast alternating terms
of Eq.~(\ref{eq6}) since they are proportional to $\xi (x)$ and
therefore are smaller than the term proportional to $g (x)$. Next, we
note that the terms included into the first pair of brackets in
Eq.~(\ref{eq7}) cancel each other independently on the terms included
into the second pair of brackets in Eq.~(\ref{eq7}) as these two type
of terms have very different length scales $l$ and $\Lambda_J$ and
$l\ll\Lambda_J$.\cite{Lan} The same reasoning is applicable to the
terms included into the second pair of brackets in Eq.~(\ref{eq7}). As
a result we obtain the following two equations for $\xi (x)$ and
$\psi (x)$\cite{RGM}
% #8,9
\begin{eqnarray}
& \Lambda_J^2\xi ''=g(x)\sin\psi,\label{eq8}\\
& \Lambda_J^2\psi''-\sin\psi-
\langle g(x)\xi(x)\rangle\cos\psi=0.\label{eq9}
\end{eqnarray}
It is worth to note that we obtain the {\it two} functions $\psi (x)$
and $\xi (x)$ from {\it one} equation (\ref{eq7}) as only two type of
terms with different typical length scales $l$ and $\Lambda_J$ appear
in Eq.~(\ref{eq6}). If $g(x)$ would have a wide range of typical
length scales the above separation would not be possible.
\par
Introducing the Fourier transform of $g(x)$ as
% #10
\begin{equation}
g(x)=\sum_{-\infty}^\infty g_ke^{ikx}
\label{eq10}
\end{equation}
we find the solution of Eq.~(\ref{eq8}) in the form
% #11
\begin{equation}
\xi (x)=-{\sin\psi\over\Lambda_J^2}\,
\sum_{-\infty}^\infty {g_ke^{ikx}\over k^2}=-\xi_g(x)\,\sin\psi,
\label{eq11}
\end{equation}
where the sums in Eqs.~(\ref{eq10}) and (\ref{eq11}) are over the wave
vectors $k=2\pi n/{\cal L}$, ${\cal L}$ is the length of the junction,
and $n$ is an integer. It is worth mentioning that the function
$\xi_g (x)$ is defined only by the alternating components of the
critical current density $j_c(x)$. Also while deriving Eq.~(\ref{eq11})
we ignored the spatial dependence of $\sin\psi$. This can be done since
on the length-scale $l$ the variation of the smooth function
$\sin\psi (x)$ is of order $l/\Lambda_J\ll 1$. The alternating part of
the critical current density has the typical wave numbers $k\sim 1/l$.
Therefore, using Eq.~(\ref{eq11}) we estimate $\xi (x)$ as
% #12
\begin{equation}
\xi (x)\sim -\sin\psi{l^2\over\Lambda_J^2}\,g(x).
\label{eq12}
\end{equation}
It follows from this estimate that the typical values of the phase
difference $\xi (x)$ are small $(\langle|\xi (x)|\rangle\ll 1)$ if
% #13
\begin{equation}
\langle |g(x)|\rangle\ll {\Lambda_J^2\over l^2}.
\label{eq13}
\end{equation}
\par
Next, using Eq.~(\ref{eq11}) we rewrite Eq.~(\ref{eq9}) for the smooth
phase shift $\psi (x)$ in the form
% #14
\begin{equation}
\Lambda_J^2\psi''-\sin\psi+\gamma\sin\psi\cos\psi=0,
\label{eq14}
\end{equation}
where
% #15
\begin{equation}
\gamma=\langle g(x)\xi_g(x)\rangle=
-{1\over \sin\psi}\,\langle g(x)\xi(x)\rangle
\label{eq15}
\end{equation}
is a constant. A similar derivation for the current density across the
tunnel junction results in
% #16
\begin{equation}
j(x)=\langle j_c\rangle\sin\psi (1 -\gamma\cos\psi).
\label{eq16}
\end{equation}
\par
The magnetic field $B_s(x)$ generated by the alternating component of
the critical current $\langle j_c\rangle\,g(x)$ is given by
% #17
\begin{equation}
B_s={4\pi\over c}\,\langle j_c\rangle\int g(x)\,dx=
-{\phi_0\over 4\pi\lambda}\,{d\xi_g\over dx}
\label{eq17}
\end{equation}
and the averaged field $\langle B_s(x)\rangle =0$. The alternating
magnetic flux $\phi_s$ produced by the field $B_s$ is equal to
% #18
\begin{equation}
\phi_s=-{\phi_0\over 2\pi}\,\xi_g.
\label{eq18}
\end{equation}
Combining Eqs.~(\ref{eq15}) and (\ref{eq17}) we find for $\gamma$ the
formula
% #19
\begin{equation}
\gamma={c\lambda\over\phi_0\langle j_c\rangle }\,\langle B_s^2\rangle=
{\langle B_s^2\rangle\over\langle B_J^2\rangle},
\label{eq19}
\end{equation}
where we introduce a characteristic magnetic field
% #20
\begin{equation}
B_J={4\pi\over c}\,\langle j_c\rangle\Lambda_J.
\label{eq20}
\end{equation}
It follows from equation (\ref{eq19}) that $\gamma$ is a positive
constant which can be estimated as
% #21
\begin{equation}
\gamma\sim {l^2\over\Lambda_J^2}\,\langle g^2\rangle.
\label{eq21}
\end{equation}
\par
The energy of a Josephson junction $\cal E$ takes the form\cite{Bar}
${\cal E}={\cal E}_0+{\cal E}_{\varphi}$, where ${\cal E}_0$ is
independent on $\varphi (x)$ and
% #22
\begin{equation}
{\cal E}_{\varphi}={\langle j_c\rangle\over 2e}\,\int dx
\Bigl\{{1\over 2}\,\Lambda_J^2\varphi'^2-[1+g(x)]\cos\varphi\Bigr\}.
\label{eq22}
\end{equation}
Using Eqs.~(\ref{eq8}), (\ref{eq11}), and the definition of $\gamma$,
we obtain the energy ${\cal E}_{\varphi}$ in terms of the smooth phase
shift $\psi (x)$
% #23
\begin{equation}
{\cal E}_{\varphi}={\hbar\langle j_c\rangle\over 4e}\int dx
\Bigl\{\Lambda_J^2\psi'^2-2\cos\psi
-\gamma\sin^2\psi\Bigr\}.
\label{eq23}
\end{equation}
Note, that solutions $\psi (x)$ of Eq.~(\ref{eq14}) correspond to the
minima and to the maxima of the energy functional (\ref{eq23}).
\par
\subsection{Non-quantized self-generated flux}
Let us apply Eqs.~(\ref{eq8}), (\ref{eq9}), and (\ref{eq15}) to a
consider the stationary states of a Josephson junction with a certain
length ${\cal L}\gg \Lambda_J$ in an absence of applied magnetic
field. In this case the average flux inside the junction is zero and
thus an alternating self-generated flux $\phi_s (x)$ appears
simultaneously with a certain phase $\psi ={\rm const}$ as it follows
from Eqs.~(\ref{eq11}) and (\ref{eq18}).
\par
In the stationary state with $\psi ={\rm const}$ the values of $\psi$
are determined by Eq.~(\ref{eq14}) which takes the form
% #24
\begin{equation}
\sin\psi\,(1-\gamma\cos\psi)=0.
\label{eq24}
\end{equation}
Note, that this equation means also that the current density $j(x)$
across the tunnel junction is equal to zero.
\par
In the case $\gamma <1$ equation (\ref{eq24}) has two solutions,
namely, $\psi =0$ and $\psi=\pi$ and thus, as follows from
Eq.~(\ref{eq11}), there is no self-generated flux. It is also worth
mentioning that the energy of a Josephson junction ${\cal E}$ has a
minimum for $\psi =0$ and maximum for $\psi =\pi$.
\par
In the case $\gamma\ge 1$ there are four solutions of equation
(\ref{eq24}), namely, $\psi =-\psi_{\gamma},\,0,\,\psi_{\gamma},\,\pi$,
where
% #25
\begin{equation}
\psi_{\gamma}=\arccos\,(1/\gamma).
\label{eq25}
\end{equation}
The energy ${\cal E}[\psi(x)]$ has a minimum for
$\psi=\pm\,\psi_{\gamma}$ and a maximum for $\psi =0,\pi$. The
self-generated flux
% #26
\begin{equation}
\phi_s=-{\phi_0 \xi\over 2\pi}=-{\phi_0\xi_g\over 2\pi}\sin\psi_\gamma=
\mp\,\phi_0{\xi_g\over 2\pi}\,{\sqrt{\gamma^2-1}\over\gamma}
\label{eq26}
\end{equation}
arises in the two stationary states with $\psi=\pm\,\psi_{\gamma}$,
each of these states corresponds to a minimum energy $\cal E$. The
assumption $\langle\xi (x)\rangle\ll 1$ restricts the value of
$\gamma$. However, it follows from Eqs.~(\ref{eq13}) and (\ref{eq21})
that $\langle\xi (x)\rangle\ll 1$ and $\gamma >1$ hold simultaneously
only if
% #27
\begin{equation}
{\Lambda_J\over l}<\langle |g(x)|\rangle\ll{\Lambda_J^2\over l^2}.
\label{eq27}
\end{equation}
Using equations (\ref{eq12}) and (\ref{eq26}) we estimate
$|\phi_s(x)|$ as
% #28
\begin{equation}
|\phi_s(x)|\sim\phi_0\,{\sqrt{\gamma^2-1}\over\gamma}\,
{l^2\over\Lambda_J^2}\,|g(x)|\ll\phi_0.
\label{eq28}
\end{equation}
\par
The above results hold only for a periodic critical current density
$j_c(x)$ and the predicted self-generated flux $\phi_s(x)$ has a
typical amplitude of variations which is much less than the one
observed experimentally.\cite{JMa}
\par
%
%%%%%%%%%%Non-periodic alternating critical current density%%%%%%%%%%%
%
\section{Non-periodic alternating critical current density}
The above analytical approach to the problem of a self-generated flux
in a non-uniform Josephson junction is based on an assumption that the
critical current density $j_c(x)$ is an alternating periodic function.
This model allows for analytical calculation and provides a reasonable
preliminary insight into the properties of an idealized Josephson
junction with an alternating $j_c(x)$. At the same time this simple
model fails for a quantitative description of any real system with a
certain randomness of the spatial distribution of Josephson critical
current density $j_c(x)$. In this section we generalize the above
approach assuming that the alternating critical current density
$j_c(x)$ is almost periodic, {\it i.e.}, we assume that there is a
typical length of interchange of sign of $j_c(x)$ which is distributed
randomly near some mean value $l$.
\par
In the case of an almost periodic $j_c(x)$ we can not apply the
two-scale perturbation theory in the same way as we did it in the
previous section. Indeed, an arbitrary solution of Eq.~(\ref{eq8})
takes the form
% #29
\begin{equation}
\xi(x)={\sin\psi\over\Lambda_J^2}\,G(x),
\label{eq29}
\end{equation}
where
% #30
\begin{equation}
G(x)=\int_{a}^x\!\! dx'\!\! \int_{a'}^{x'} \!\! dx''\,g(x''),
\label{eq30}
\end{equation}
and the integration constants $a$ and $a'$ are defined by the boundary
conditions. The random function $G(x)$ increases with an increase of
the integration interval. In general, the value of $|G(x)|$ can become
arbitrarily large if the length of the tunnel junction $\cal L$ becomes
large enough. This is in a contradiction with our main assumption that
the phase $\xi (x)$ is a small and fast varying component of the total
phase difference $\varphi (x)$. To solve this contradiction we write
$\xi (x)$ as
% #31
\begin{equation}
\xi(x)={\sin\psi\over \Lambda^2_J}\left[ G(x)-G_a(x)\right].
\label{eq31}
\end{equation}
The function $G_a(x)$ is a smoothing average of $G(x)$ over an interval
with a certain length ${\cal L}_a$, where $l<{\cal L}_a\ll {\cal L}$.
\par
The procedure of filtering out the smooth part of $G(x)$ is especially
evident if we use Fourier series for $g(x)$ and $G(x)$. Introducing
Fourier transform for $g(x)$ as
% #32
\begin{equation}
g(x)=\sum_{-\infty}^\infty g_k\,e^{ikx}
\label{eq32}
\end{equation}
we find Fourier series for the function $G(x)$ in the form
% #33
\begin{equation}
G(x)=-\sum_{-\infty}^\infty {g_k\over k^2}\,e^{ikx},
\label{eq33}
\end{equation}
where the sums in Eqs.~(\ref{eq32}) and (\ref{eq33}) are over the wave
vectors $k=2\pi n/{\cal L}$ and $n$ is an integer. The smooth part of
the function $G(x)$ can be obtained by extracting the fast Fourier
harmonics, {\it i.e.}, by extracting from the sum (\ref{eq33}) terms
with wave vectors $|k|>k_a\sim 2\pi/{\cal L}_a$. As a result we find
for $G_a(x)$ and $\xi(x)$ the series
% #34,35
\begin{eqnarray}
G_a(x)&=&-\sum_{-k_a}^{k_a} {g_k\over k^2}\,e^{ikx},
\label{eq34}\\
\xi(x)&=&{\sin\psi\over\Lambda_J^2}\,
\Biggl (\sum_{-\infty}^{-k_a} + \sum_{k_a}^{\infty} \Biggr )
\Biggl [{g_k\over k^2}\,e^{ikx}\Biggr ].
\label{eq35}
\end{eqnarray}
\par
The small and fast alternating part $\xi (x)$ of the phase difference
$\varphi (x)$ is thus defined by Eq.~(\ref{eq31}). This equation is a
straightforward generalization of the two-scale perturbation theory
approach to a real case of an almost periodic $g(x)$.
\par
Next, we use Eq.~(\ref{eq31}) to derive an equation for the smooth
part $\psi (x)$ of the phase $\varphi (x)$. {\it First}, combining
Eqs.~(\ref{eq31}), (\ref{eq7}), and (\ref{eq30}) we arrive to the
relation
% #36
\begin{equation}
\Lambda^2_J\psi''-[1+G_a''(x)]\sin\psi-g(x)\xi(x)\sin\psi=0.
\label{eq36}
\end{equation}
{\it Second}, we average Eq.~(\ref{eq36}) over an interval with a
certain length $L$ assuming that $l\ll L\ll {\cal L}_a$. This averaging
results in an equation describing the phase $\psi(x)$
% #37
\begin{equation}
\Lambda^2_J\psi''-[1+G_a''(x)]\sin\psi +
\gamma (x)\sin\psi\cos\psi=0,
\label{eq37}
\end{equation}
where $\gamma (x)$ is defined by Eq.~(\ref{eq15}). It is worth
mentioning that equation (\ref{eq37}) differs from the analogues
equation (\ref{eq9}) by an additional term $G_a''(x)$ in the
coefficient before $\sin\psi$ and by the fact that the parameter
$\gamma =\gamma (x)$ is a function of the coordinate $x$ along the
junction.
\par
The coefficient $1+G_a''(x)$ is defined by magnetic field $B_c(x)$
which would be
generated by a current with the density $j_c(x)$. Indeed, using
Eqs.~(\ref{eq1}), (\ref{eq30}), and Maxwell's equation
$dB_c/dx=4\pi j_c/c$ we obtain for $G(x)$
% #38
\begin{eqnarray}
G(x)&=&{c\over 4\pi \langle j_c\rangle}\,\int dx'\,[B_c(x')-B_a(x')]=
\nonumber\\
&=&{2\pi\Lambda_J^2\over\phi_0}\,[\phi_c(x)-\phi_a(x)],
\label{eq38}
\end{eqnarray}
where the magnetic field $B_a(x)$ would be generated by a constant
current density $\langle j_c\rangle$, {\it i.e.}, $dB_a/dx=4\pi
\langle j_c\rangle/c$, and $\phi_c(x)$ and $\phi_a(x)$ are the fluxes
of the fields $B_c(x)$ and $B_a(x)$. It follows now from
Eq.~(\ref{eq38}) that
% #39
\begin{equation}
1+G_a''(x)={2\pi\Lambda^2_J\over \phi_0}\,\langle\phi_c(x)\rangle''.
\label{eq39}
\end{equation}
\par
As shown above the value of the parameter $\gamma$ is determining the
existence or absence of the self-generated flux. Using Eq.~(\ref{eq8})
and the definition of $\gamma (x)$ given by Eq.~(\ref{eq15}) we obtain
for $\gamma (x)$ an expression
% #40
\begin{equation}
\gamma (x)=-{\Lambda^2_J\over\sin^2\psi}\,\langle \xi''\xi\rangle =
{\Lambda^2_J\over\sin^2\psi}\,\langle {\xi'}^2\rangle >0
\label{eq40}
\end{equation}
demonstrating that the condition $\gamma (x)>0$ holds also in the case
of an almost periodic critical current density.
\par
\subsection{Self-generated flux in a stationary state}
Let us now consider stationary solutions for the smooth part $\psi (x)$
of the phase difference $\varphi (x)$ qualitatively. Assume {\it first}
that there are sufficiently large intervals with lengths $L_i\gg\Lambda_J$,
where the function $\psi(x)$ is constant or varies with a typical
space-scale of order ${\cal L}_i\gg\Lambda_J$. In this case
equation (\ref{eq37}) reduces to
% #41
\begin{equation}
\big[1+G_a''(x)-\gamma(x)\cos\psi\big]\sin\psi=0.
\label{eq41}
\end{equation}
\par
%%%%%%%%%% FIGURE #1 %%%%%%%%%%
\epsfclipon
\vskip -.75\baselineskip
\begin{figure}
\epsfxsize =\hsize
\centerline{\epsfbox {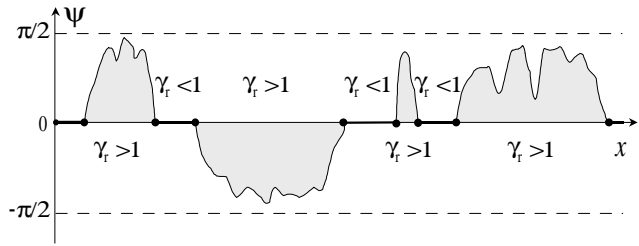}}
\vskip \baselineskip
\caption{Scheme of the phase difference distribution $\psi(x)$ as an
alternation of two types of solutions corresponding to different values
of $\gamma_r$.}
\label{fig_1}
\end{figure}
%%%%%%%%%% FIGURE #1 %%%%%%%%%%
\par
This equation is similar to Eq.~(\ref{eq24}) which we derived for the
case of a periodic critical current density $j_c(x)$ and has different
solutions depending on the value of the parameter
% #42
\begin{equation}
\gamma_r (x)={\gamma(x)\over 1+G_a''(x)}. \label{eq42}
\end{equation}
\par
In the regions with $\gamma_r (x)<1$ equation (\ref{eq41}) has two
solutions $\psi =0$ and $\psi =\pi$ and therefore as it follows from
Eq.~(\ref{eq29}) there is no self-generated flux in these regions.
\par
The energy of a Josephson junction ${\cal E}_\varphi$ given by
Eq.~(\ref{eq22}) can be written in terms of the smooth part of the
phase difference $\psi (x)$. In the case of an almost periodic
critical current density $j_c(x)$ this equation reads
% #43
\begin{eqnarray}
& &{\cal E}_{\varphi}={\hbar\langle j_c\rangle\over 4e}\times
\nonumber \\
& &\int dx\Bigl\{\Lambda_J^2\psi'^2-2\cos\psi [1+G_a''(x)] -
\gamma (x)\sin^2\psi\Bigr\}
\label{eq43}
\end{eqnarray}
and if $\gamma_r (x)<1$, then the energy ${\cal E}_{\varphi}$
has a minimum for $\psi =0$ and a maximum for $\psi =\pi$.
\par
In the regions where the function $\gamma_r(x)>1$ equation (\ref{eq41})
has four solutions $\psi =-\psi_\gamma (x),\,0,\,\psi_\gamma (x),\,
\pi$, with
% #44
\begin{equation}
\psi_{\gamma} (x)=\arccos\biggl[{1\over\gamma_r(x)}\biggr]=
\arccos\biggl[{1+G_a''(x)\over\gamma (x)}\biggr].
\label{eq44}
\end{equation}
The energy ${\cal E}_{\varphi}$ has a minimum for $\psi
=\pm\,\psi_\gamma (x)$ and a maximum for $\psi =0,\,\pi$. The
self-generated flux is thus non-zero in the regions with $\gamma_r>1$.
This flux has a fast and a smoothly varying parts defined by $\xi (x)$
and $\psi_\gamma (x)$.
\par
%%%%%%%%%% FIGURE #2 %%%%%%%%%%
\epsfclipon
\begin{figure}
\epsfysize 0.75\hsize
\centerline{\epsfbox {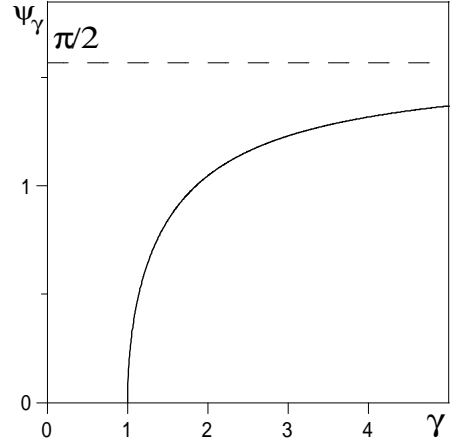}}
\caption{The  dependence of $\psi_\gamma$ on $\gamma$.}
\label{fig_2}
\end{figure}
%%%%%%%%%% FIGURE #2 %%%%%%%%%%
\par
The randomness of the function $g(x)$ causes variation of
$\gamma_r (x)$ along the junction. As a result of this variation
intervals with $\gamma_r (x)>1$ are interlaced with intervals with
$\gamma_r (x)\leq 1$. As it was mentioned above, in the case of
$\gamma_r >1$ the energy of the Josephson junction ${\cal E}_\varphi$
has a minimum for $\psi =\pm\,\psi_{\gamma}(x)$ and a maximum for
$\psi =0,\,\pi$. When the value of $\gamma_r (x)$ changes from
$\gamma_r (x)>1$ to $\gamma_r (x)\leq 1$ the energy
${\cal E}_\varphi$ still has a maximum if $\psi=\pi$, but a state with
$\psi=0$ becomes a state with a minimum energy.
\par
The above results provide a qualitative description of experimentally
observable flux distribution along a Josephson junction with an almost
periodic alternating critical current density. This flux distribution
spatially averaged by the measurement tools is defined by the function
$\psi(x)$ (see Fig.~\ref{fig_1}). Inside the intervals with
$\gamma_r(x)>1$ the phase $\psi(x)$ tends to one of the solutions
$\pm\,\psi_{\gamma}(x)$. The profile of the function $\psi(x)$
correlates with the profile of $\psi_{\gamma}(x)$, though does not
coincide with it exactly because the solution
$\psi(x)=\psi_{\gamma}(x)$ was obtained under assumption $\psi''=0$,
which does not hold exactly for the intervals $\gamma_r(x)>1$. The
smooth part of the phase difference inside the intervals with
$\gamma_r(x)\leq 1$ is $\psi = 0$ which is consistent with the
assumption $\psi''=0$.
\par
The value of $\psi_\gamma$ increases quite fast with an increase of
the parameter $\gamma$ (see Fig.~\ref{fig_2}). In particular, for
$\gamma =2$ the value of $\psi_\gamma$ is already about 0.75 of its
maximum value $\pi /2$. This means that for most of the experimentally
observable peaks of the self-generated flux the values of $\psi$ will
be close to $\pi /2$ which corresponds to a magnetic flux $\phi_0 /4$.
In some places of the junction the phase $\psi$ changes from
$-\psi_\gamma$ to $\psi_\gamma$. The flux localized in this area of
the junction will be close to $\phi_0/2$.
\par
%
%%%%%%%%%%%Numerical Simulations%%%%%%%%%%
%
\section{Numerical simulations}
\subsection{The finite difference scheme}
To study the self-generation of magnetic flux in a tunnel junction
with an alternating critical current density numerically we introduced
time dependence into the main equation (\ref{eq4})
% #45
\begin{equation}
\ddot{\varphi} +\alpha\dot{\varphi}-\Lambda_J^2\psi''
+[1+g(x)]\sin\varphi=0,
\label{eq45}
\end{equation}
where $\alpha\sim 1$ is a decay constant. This approach allows to study
both dynamics and statics of the system.
\par
The term $\alpha\dot{\varphi}$ introduces dissipation. As a result of
this dissipation the relaxation of the system ends up in one of the
stable stationary states described by a certain solution of the static
equation (\ref{eq4}). Moreover, for a given distribution of the
critical current density $j_c(x)$ we obtained different static
solutions when we start the numeric simulation from different initial
states. We compare and classify these solutions based on the features
of the function $j_c(x)$. Indeed, this function essentially describes
the pinning properties of the junction. Therefore a variety of initial
conditions can converge to a similar flux pinning pattern.
\par
To solve Eq.~(\ref{eq45}) numerically we use the finite difference
scheme.\cite{Abl} We adopted this method to our case and checked
stability and convergency of the obtained solutions. As a result we
arrived to the following scheme
% #46,47,48,49
\begin{eqnarray}
&\varphi&\rightarrow{\varphi_{m+1}^n+\varphi_{m-1}^n \over 2}
\equiv\tilde{\varphi}_m^n\\
&\dot{\varphi}&\rightarrow
{\tilde{\varphi}_m^n-\varphi_m^{n-1} \over\tau}\\
&\ddot{\varphi}&\rightarrow{\varphi_m^{n+1}+\varphi_m^{n-1}-2
\tilde\varphi_m^n \over\tau^2}\\
&\varphi''&\rightarrow{\varphi_{m+1}^n+\varphi_{m-1}^n-2
\varphi_m^n \over h^2},
\end{eqnarray}
where $f^n_m=f(x_m,t_n)$, $\tau$ and $h$ are steps along $t$ and $x$
correspondingly. Next, we choose units providing $\Lambda_J=1$ and set
$h$=$\tau$. As a result we arrive to the following finite difference
scheme
% #50
\begin{eqnarray}
\varphi_m^{n+1}=&-&(1- \alpha\tau)\,\varphi_m^{n-1}+
(2-\alpha\tau)\,\tilde{\varphi}_m^n
\nonumber\\
&-&\tau^2(1-g_m)\sin\tilde{\varphi}_m^n.
\label{eq50}
\end{eqnarray}
\par
\subsection{Stationary solutions}
Initially a certain random function $g(x)$ is generated for an interval
with a length $L$ with a given values of $l$ and $\delta l$ (a typical
length-scale of the function $g(x)$ and its dispersion), $g$ and
$\delta g$ (amplitude of the function $g(x)$ and its dispersion). This
allows to calculate the function $\gamma_r (x)$ for the whole
interval. An initial state $\varphi_0(x)$ is prepared as a random or
some specific function. Finally the dynamical rules (\ref{eq50}) are
applied to the initial state iteratively until a stationary state is
established.
\par
\begin{figure}
\epsfxsize \hsize \centerline{\epsfbox {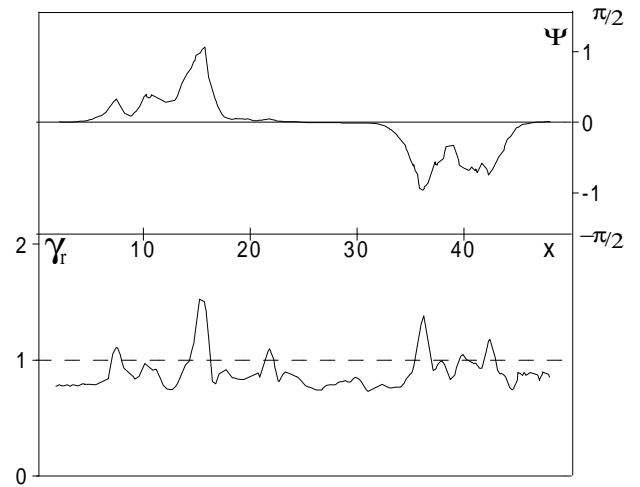}}
\caption{Profiles of $\gamma_r(x)$ and the corresponding $\psi(x)$.}
\label{fig_3}
\end{figure}

In Fig.~\ref{fig_3} we show one of the stationary solutions obtained by
a numerical simulation and the function $\gamma_r (x)$ calculated for
this solution. It is clearly seen from Fig.~\ref{fig_3} that
$\varphi(x)$ arises at the places where $\gamma_r (x)$ exceeds $1$.
Heights of the peaks are less than $\pi/2$, and thus the corresponding
magnetic flux amplitudes are less than $\phi_0/4$.
\par
%%%%%%%%%% FIGURE #3 %%%%%%%%%%
\epsfclipon
%%%%%%%%%% FIGURE #3 %%%%%%%%%%
\par
In general a different initial state of the same sample, {\it i.e.},
for the same function $\gamma_r (x)$, generates a different stationary
state. Our numerical simulations show that these different states
differ only by sign of some peaks of $\varphi(x)$, but the shapes and
locations stay unchanged.
\par
We have compared our results with the experimental data.\cite{JMa} The
typical amplitude of the flux variations measured by a SQUID pickup
loop with a size of several $\Lambda_J$ is about $0.25$ of $\phi_0$
with rare narrow picks with an amplitude about $0.5\,\phi_0$ which is
in a good agreement with our calculations.
\par
\section{SUMMARY}
We treated a Josephson junction with an alternating critical current
density $j_c(x)$ as a model for considering electromagnetic properties
of grain boundaries in YBa$_2$Cu$_3$O$_{7-x}$ superconducting films.
The study is mainly focused on a specific case of an almost
periodically alternating function $j_c(x)$. We demonstrated both
analytically and numerically that under certain conditions a
self-generated flux pattern arise for this type of spatial
distribution of the critical current density $j_c(x)$. The obtained
flux pattern with two types of interlacing flux domains is similar to
the one which was recently observed experimentally in
YBa$_2$Cu$_3$O$_{7-x}$ superconducting films in the absence of an
external magnetic field. The typical amplitude for the magnetic flux
peaks is of order $\phi_0/4$ and $\phi_0/2$ and the typical distance
between the peaks depends on the spatial distribution of $j_c(x)$.
\par
\acknowledgments
One of us (RGM) is grateful to E.H. Brandt, J.R. Clem,
H. Hilgenkamp, V.G. Kogan, and J. Mannhart for useful and stimulating
discussions. This research was supported in part by grant No.~96-00048
from the United States -- Israel Binational Science Foundation (BSF),
Jerusalem, Israel.
\par
%%%%%%%%%%%%%%%%%%%%%%%%%%%%%%%%%%%%%%%%%%%%%%%%%

%
\end{multicols}

\begin{references}
% 1
\bibitem{Lin} P. Chaudhari and S.Y. Lin, \prl{\bf 72}, 1084 (1994).
% 2
\bibitem{Tsu} C.C. Tsuei, J.R. Kirtley, C.C. Chi, Lock See Yu-Jhanes,
A. Gupta, T. Shaw, J.Z. Sun, and M.B. Ketchen, \prl{\bf 73}, 593 (1994).
% 3
\bibitem{Mil} J.H. Miller, Q.Y. Ying, Z.G. Zou, N.Q. Fan, J.H. Xu,
M.F. Davis, and J.C. Wolfe, \prl{\bf 74}, 2347 (1995).
% 4
\bibitem{Cha} P. Chaudhari, D. Dimos, and J. Mannhart, IBM J. Res.
Develop. {\bf 33}, 299 (1989).
% 5
\bibitem{Gro} R. Gross, P. Chaudhari, D. Dimos, A. Gupta, and G. Koren,
\prl {\bf 64}, 228 (1990).
% 6
\bibitem{Iva} Z.G. Ivanov, J.A. Alarco, T. Claeson, P.-\AA. Nilsson, E. Olsson,
H.K. Olsson, E.A. Stepantsov, A.Ya. Tzalenchuk, and D. Winkler, in
{\it Proceedings of the Beijing International Conference on High-Temperature
Superconductivity (BHTSC '92)}, edited by Z.Z. Gan, S.S. Xie, and Z.X. Zhao
(World Scientific, Singapore, 1992), p. 722.
% 7
\bibitem{Che} N.G.~Chew {\it et al.}, Appl. Phys. Lett. {\bf 60}, 1516
(1992).
% 8
\bibitem{Hum} R.G. Humphreys, J.S. Satchell, S.W. Goodyear, N.G. Chew, M.N. Keene,
J.A. Edwards, C.P. Barret, N.J. Exon, and K. Lander, in {\it Proceedings
of 2nd Workshop on HTS Applications and New Materials}, edited by
D.H.A. Blank (University of Twente, Enschede, 1995), p.16.
% 9
\bibitem{Cop} C.A. Copetti, F. R\"{u}ders, B. Oelze, Ch. Buchal, B. Kabius,
and J.W. Seo, Physica C {\bf 253}, 63 (1995).
% 10
\bibitem{Man} H. Hilgenkamp, J. Mannhart, and B. Mayer, \prb {\bf 53},
14586 (1996).
% 11
\bibitem{Min} R.G. Mints and V.G. Kogan, \prb {\bf 55}, R8682 (1997).
% 12
\bibitem{Jia} C.L. Jia, B. Kabius, K. Urban, K. Herrmann, J. Schubert,
W. Zander, and A.I. Braginski, Physica C {\bf 196}, 211 (1992).
% 13
\bibitem{Ros} S.J. Rosner, K. Char, and G. Zaharchuk, Appl. Phys. Lett.
{\bf 60}, 1010 (1992).
% 14
\bibitem{Tra} C. Tr\ae holt, J.G. Wen, H.W. Zandbergen, Y. Shen, and
J.W.M. Hilgenkamp, Physica C {\bf 230}, 425 (1994).
% 15
\bibitem{Seo} J.W. Seo, B. Kabius, U. D\"{a}hne, A. Scholen, and K. Urban,
Physica C {\bf 245}, 25 (1995).
% 16
\bibitem{DJM} D.J. Miller, T.A. Roberts, J.H. Kang, J. Talvacchio,
D.B. Buchholz, and R.P.H. Chang, Appl. Phys. Lett. {\bf 66}, 2561
(1995).
% 17
\bibitem{Wol} D.A. Wollman, D.J. Van Harlingen, D.J. Lee, W.C. Lee,
D.M. Ginsberg, and A.J. Legget, \prl {\bf 71}, 2134 (1993).
% 18
\bibitem{Bra} D.A. Brawner and H.R. Ott, \prb {\bf 50}, 6530 (1994).
% 19
\bibitem{Igu} I. Iguchi and Z. Wen, \prb {\bf 49}, 12388 (1994).
% 20
\bibitem{Van} D.J. Van Harlingen, \rmp {\bf 67}, 515 (1995).
% 21
\bibitem{Ish} Y. Ishimaru, J. Wen, K. Hayashi, Y. Enomoto, and N. Koshizuka,
Jpn. J. Appl. Phys. {\bf 34}, L1532 (1995).
% 22
\bibitem{Mul} K.A. M\"{u}ller, Nature {\bf 377}, 133 (1995).
% 23
\bibitem{JMa} J. Mannhart, H. Hilgenkamp, B. Mayer, Ch. Gerber,
J.R. Kirtley, K.A. Moler, and M. Sigrist, \prl {\bf 77}, 2782 (1996).
% 24
\bibitem{RGM} R.G. Mints, \prb {\bf 55}, R8682 (1997).
% 25
\bibitem{Lan} L.D. Landau and E.M. Lifshitz, {\it Mechanics} (Pergamon
Press, Oxford, 1994), p. 93.)
% 26
\bibitem{Bar} A. Barone and G. Paterno, {\it Physics and Applications of
the Josephson Effect}, Wiley, New York, 1982).
% 27
\bibitem{Abl} M.J. Ablowitz, M.D. Kruskal, J.F.Ladic, {\it Solitary
wave collisions}, SIAM J. Appl. Math., {\bf 36}, 428 (1997).
%
\end{references}
\end {document}